\documentclass[12pt,preprint]{aastex}
\begin{document}

\parindent=1.0cm

\title
{Ghosts in the Attic: Mapping the Stellar Content of the S0 Galaxy NGC 5102
\altaffilmark{1}\altaffilmark{2}}

\author{T. J. Davidge}

\affil{Herzberg Institute of Astrophysics,
\\National Research Council of Canada, 5071 West Saanich Road,
\\Victoria, B.C. Canada V9E 2E7\\ {\it email: tim.davidge@nrc.ca}}

\altaffiltext{1}{Based on observations obtained with the Megaprime/MegaCam,
a joint project of the CFHT and CEA/DAPNIA, at the Canada-France-Hawaii Telescope,
which is operated by the National Research Council of Canada, the Institut National
des Sciences de l'Univers of the Centre National de la Recherche Scientifique of
France, and the University of Hawaii.}

\altaffiltext{2}{Based on observations obtained with WIRCam, a joint project of the CFHT, 
Taiwan, Korea, Canada and France, at the Canada-France-Hawaii Telescope,
which is operated by the National Research Council of Canada, the Institut National
des Sciences de l'Univers of the Centre National de la Recherche Scientifique of
France, and the University of Hawaii.}

\begin{abstract}

	The spatial distribution of stars in the 
nearby S0 galaxy NGC 5102 is investigated using 
images obtained with WIRCam and MegaCam on the Canada-France-Hawaii Telescope. 
With the exception of gaps between detector elements, 
the entire galaxy is surveyed in $r'$ and $i'$, while the $J$ and $Ks$ data extend out to 
R$_{GC} \approx 6$ kpc, which corresponds to almost 7 disk scale 
lengths. A modest population of main sequence (MS) stars with M$_V < -3.5$ 
and ages $\approx 70$ Myr are detected throughout the disk, with the majority 
located in the southern half of the galaxy. The stellar disk in the northern half of 
the galaxy is warped, following structure that is also seen in HI. 
Objects with photometric properties that are consistent with those of bright AGB stars 
are seen throughout the disk, and the ratio of C stars to bright M giants 
is consistent with an overall increase in the star formation rate within the past 1 Gyr. 
Star-forming activity during the interval $0.1 - 2$ Gyr was more centrally concentrated 
than during the past $\approx 100$ Myr. The structure of the 
disk changes near R$_{GC} \approx 5$ kpc (5.5 disk scale lengths), 
in the sense that the radial surface density profile defined by red supergiants (RSGs) and 
bright AGB stars levels off at larger radii. RSGs and bright AGB stars are traced out 
to a radius of 14 kpc (15.6 scale lengths) along the southern portion of the major 
axis, while a tentative detection is also made of bright AGB stars at 
a projected distance of $\approx 16$ kpc along the south east minor axis. 
A large clump of AGB stars that subtends $\approx 1$ arcmin 
is identified to the west of the galaxy center. It is argued that this is the remnant 
of a companion galaxy that triggered past episodes of elevated star-forming activity. 

\end{abstract}

\keywords{galaxies: individual (NGC 5102) - galaxies: elliptical and S0, cD
- galaxies: stellar content - galaxies: evolution}

\section{INTRODUCTION}

	The stellar contents of nearby galaxies constitute a fossil record that can be 
used to gain insight into the processes that shape galaxy evolution and define basic 
global properties, such as morphology. There are five galaxy groups within $\approx 4$ Mpc 
(classically referred to as the Local, Sculptor, M81, Centaurus, and Maffei Groups), 
which in turn contain sub-systems centered on the largest members 
(e.g. Karachentsev 2005). The brightest members of the nearest groups show a diverse range 
of properties, although a common feature shared by the most massive members of the 
Local, M81, and Centaurus groups is that they show signs of having experienced 
cosmologically recent interactions. This commonality underlines the important role that 
local environment and galaxy-galaxy interactions play in the evolution of even 
relatively commonplace galaxies.

	Located at a distance of only 3.2 Mpc, 
the S0 galaxy NGC 5102 is a key laboratory for probing the evolution 
of gas-poor disks. There is only modest star-forming activity in NGC 5102 today, and 
this is restricted to the south east quadrant of the disk, where HII regions 
are found (van den Bergh 1976, McMillan, Ciardullo, \& Jacoby 1994). 
Davidge (2008a) found small numbers of red supergiants (RSGs) and 
blue main sequence (MS) stars in a field along the north west minor axis, 
and estimated that the integrated SFR during the past ten million years was 
0.02 M$_{\odot}$ year$^{-1}$. The moderately deep color-magnitude diagram (CMD) 
of the northern disk of NGC 5102 discussed by Karachentsev et al. (2002) 
shows that while there is a spray of blue objects, there is neither a well-defined bright 
MS nor a RSG sequence in this part of NGC 5102, again consistent with a low recent SFR. 

	There are indications that the specific SFR in NGC 5102 within the past few hundred 
Myr was much higher than it is today, at least in some parts of the galaxy. The deep Balmer 
absorption lines (Gallagher, Faber, \& Balick 1975; Rocca-Volmerange \& Guiderdoni 1987) and 
blue colors (van den Bergh 1976; Pritchet 1979; Bica 1988; Kraft et al. 2005) that are 
seen in the central regions of NGC 5102 are perhaps the most overt signatures that at least 
10\% (e.g. Serra \& Trager 2007) of the central stellar mass formed during intermediate 
epochs. Applying spectral synthesis techniques to the ultraviolet -- 
visible spectrum of the center of NGC 5102, Kraft et al. (2005) conclude that at least 
50\% of the central stellar mass in NGC 5102 formed within the past 3 Gyr. 

	There is also evidence of recent episodes of large-scale star formation in other 
parts of the galaxy. van den Bergh (1976) detected filamentary H$\alpha$ emission, which 
McMillan et al. (1994) subsequently found to be part of 
a complex web of emission features, the dominant structure of which 
is a supershell. The size and expansion velocity of the supershell 
suggest an age of $\approx 10$ Myr. Finally, the numbers of bright asymptotic 
giant branch (AGB) stars in the western portion of NGC 5102 suggests 
that $\approx 20\%$ of the total stellar mass in this field may have formed during the past 
Gyr; assuming that the stellar content of this area is representative 
of the galaxy as a whole then the galaxy-wide SFR was at least 1.4 
M$_{\odot}$ year$^{-1}$ throughout much of the past Gyr (Davidge 2008a). 

	Elevated levels of star-forming activity are usually associated with interactions, 
and so it is puzzling that NGC 5102 has no obvious close companions. The distance and 
radial velocity of NGC 5102 are consistent with it being a companion of Cen A 
(Karachentsev et al. 2002), although it is located near the outer periphery of Cen A 
satellites. Indeed, NGC 5102 is relatively isolated when compared with other galaxies 
associated with Cen A and M83 (e.g. Figure 1 of Karachentsev et al. 2002).

	While sampling relatively faint objects, the photometric 
observations discussed by Karachensev et al. (2002) and Davidge 
(2008a) cover only a small fraction of NGC 5102, missing the areas 
near the HII regions where the richest concentrations of young stars should lurk. 
These data are also restricted to visible and red wavelengths, and so are biased against 
stars that may have their photometric properties affected by line blanketing, have 
very low effective temperatures, and/or are highly obscured. These observational 
deficiencies, coupled with the evidence that there was 
much more extensive star formation throughout NGC 5102
within the past $\approx$ Gyr, provide a clear motivation 
for conducting a detailed census of the bright stellar content throughout the galaxy. 

	In the present study, observations obtained 
with the MegaCam and WIRCam imagers on the 3.6 meter Canada-France-Hawaii Telescope (CFHT) 
are used to survey the stellar content in and around NGC 5102.
The data obtained with these instruments are complementary. The brightest 
NGC 5102 stars in the MegaCam observations are young MS stars and RSGs, while those 
in the WIRCam observations are highly evolved oxygen-rich and carbon-rich AGB stars. 
Observations of highly-evolved cool stars at infrared wavelengths 
benefit from the enhanced contrast with respect to the comparatively 
blue body of unresolved stars in a galaxy, while observations of bright MS stars at visible 
wavelengths exploit the low number density of blue background galaxies in the magnitude 
range of interest, making bright MS stars potentially important probes of
structure in areas of low stellar mass density (e.g. Davidge 2008b).

	The paper is structured as follows. The observations and data 
reduction procedures are described in \S 2, while the CMDs obtained from these data are 
discussed in \S 3. The statistics of M giants and C stars in the WIRCam data are used 
to investigate the recent star-forming history of NGC 5102 in \S 4. The spatial 
distribution of stars spanning a range of ages and evolutionary states are examined in 
\S 5, while a summary and discussion of the results follows in \S 6.

\section{OBSERVATIONS \& REDUCTIONS}

\subsection{MegaCam}

	NGC 5102 was observed with MegaCam (Boulade et al. 2003) during the night of 
January 16, 2008. The detector in MegaCam is a mosaic of 36 CCDs that covers a 
$1 \times 1$ degree$^2$ field with 0.185 arcsec pixel$^{-1}$. There is a 13 arcsec gap 
between most CCDs, although there are two 70 arcsec gaps that separate 
CCD banks. The nucleus of NGC 5102, where individual stars can not be resolved, was 
centered on the detector mosaic, and thus fell in a gap between detectors.

	Four 300 sec exposures were recorded each in $r'$ and $i'$, and 
the raw images were reduced with the ELIXER pipeline. The ELIXER processing consisted of 
bias subtraction, the division by a flat-field frame, the subtraction of a fringe frame 
from the $i'$ images, and the insertion into the image header of photometric calibration 
information based on standard star observations that were made during the MegaCam run. To 
create final images, the ELIXER-processed data in each filter were aligned, median-combined, 
and trimmed to the area common to all frames. Stars in the final images 
have FWHM $\approx 1$ arcsec.

\subsection{WIRCam}

	NGC 5102 was observed with WIRCam (Puget et al. 2004) on the night of February 17 
2008. The detector in WIRCam is a mosaic of four $2048 \times 2048$ HgCdTe arrays, which 
together image $20 \times 20$ arcmin$^2$ with 0.3 arcsec pixel$^{-1}$. Gaps 
between the arrays have a width of $\approx 45$ arcsec. 

	Twenty 45 sec exposures were recorded in $J$, and 
eighty 20 sec exposures were recorded in $K_s$. 
The disk of NGC 5102 extends over many arcmin$^2$, and this complicates efforts to 
construct the calibration frames that are needed to remove 
interference fringes and thermal emission signatures. While the calibration frames could 
be constructed from observations of a separate background field, 
time constraints ruled out this approach, and so it was decided to sacrifice the 
angular coverage of the final processed image to enable the construction of the required 
calibration frames. Images were recorded with NGC 5102 
centered on each array of the mosaic in turn; this corresponds to a 10 arcmin offset 
between pointings. A four-point square dither pattern with 3 arcsec amplitude was 
also implemented at each offset position to further facilitate the suppression of stars, 
cosmic rays, and isolated bad pixels. 

	The raw WIRCam images were processed with the I'IWI pipeline, which performed 
dark subtraction and the division by flat-field frames. Calibration images to remove 
interference fringes and thermal signatures were constructed by median-combining the 
unregistered flat-fielded images in each filter, and the results were 
subtracted from the flat-fielded data. The processed images 
were then aligned, median-combined, and trimmed to the region having a common exposure time.
The region with the full exposure time covers a $10 \times 10$ arcmin area, and this is 
sufficient to sample much of the NGC 5102 disk. Stars in the final images have 
FWHM $\approx 0.9$ arcsec.

\section{THE CMDs}

\subsection{Photometric Measurements}

	The photometric measurements were made with the point-spread-function (PSF) 
fitting program ALLSTAR (Stetson \& Harris 1988), using source lists, PSFs, and preliminary 
brightnesses obtained from DAOPHOT (Stetson 1987) tasks. 
Between 50 to 100 stars were used to construct each PSF. 
The PSFs were constructed iteratively, with contaminating objects 
subtracted using progressively improved PSFs. 

	ALLSTAR computes an uncertainty, $\epsilon$, for each object that is based on the 
quality of the PSF fit, but does not account for systematic errors due to 
crowding or the tendency for the faintest stars to be located on positive noise flucuations. 
$\epsilon$ was used to cull objects with large measurement errors 
from the photometric catalogues. Objects with $\epsilon \geq 0.3$, 
which tend to be near the faint limit of the data, were rejected. 
In addition, a plot of $\epsilon$ versus magnitude 
reveals a well-defined relation, and outlying objects from the dominant 
trend were deleted. Objects that depart from the dominant trend tend to be either 
obvious blends in the crowded central regions of NGC 5102, 
galaxies, or residual cosmetic defects.

	Artificial star experiments were used to estimate completeness. 
The artificial stars were assigned $r'-i'$ and $J-K$ colors that follow 
the ridgeline of the AGB sequence in NGC 5102. The brightnesses of the artificial stars 
were measured using the same procedures employed 
for actual stars, including rejection based on $\epsilon$ 
(see above). As with actual stars, an artificial star was only considered to 
be recovered if it was detected in two filters, and so could be placed on either the 
$(i', r'-i')$ or $(K, J-K)$ CMDs.

	The completeness fraction varies with position in NGC 5102. 
The 50\% completeness fraction for stars with R$_{GC}$ between 3 and 4 kpc occurs near 
$i' = 24$, and is 0.5 magnitudes fainter for the outer regions of NGC 5102. For comparison, 
the 50\% completeness fraction occurs at $K \approx 20.5$ throughout most of the WIRCam image. 
Unlike the $i'$ data, the completeness fraction in $K$ is not as 
sensitive to location within the galaxy because of the comparatively high contrast 
between the brightest resolved stars in $K$ and the main body of unresolved stars. 
The artificial star experiments also indicate that many of the stars 
fainter than the 50\% completeness limit in $i'$ and $K$ are probably blends. 

\subsection{$(i', r'-i')$ CMDs}

	The $(i', r'-i')$ CMDs of stars in various radial intervals are shown in 
Figure 1. The distance intervals given in each panel are galactocentric radii, R$_{GC}$, 
that assume a distance modulus of 27.5 and a disk inclination of 64.4 
degrees (Davidge 2008a). The exponential disk scale length for NGC 5102 measured from 
2MASS Large Galaxy Catalogue (Jarrett et al. 2003) images is 0.9 kpc, and this is in 
excellent agreement with the scale length measured by Freeman (1970). 
While the CMDs do not go as deep as those presented by Karachentsev et al. (2002) and 
Davidge (2008a), they extend out to many disk scale lengths
and sample stars over the entire galaxy. 

	Stars evolving on the AGB form a diffuse cloud of objects with $i' > 22$ and 
$r'-i'$ between 0.2 and 1.2 in the CMDS of objects with R$_{GC} < 7$ kpc. 
It becomes difficult to identify a concentration of AGB stars in the CMDs at larger R$_{GC}$ 
due to contamination from field stars and background galaxies. Still, a modest population of 
AGB stars is present at larger R$_{GC}$ (\S 5).

	Davidge (2008a) detected blue MS stars and RSGs along the western 
arm of the NGC 5102 minor axis, and objects with $r'-i' < 0$ and $i' < 24$ 
are seen in the R$_{GC} < 5$ kpc CMDs. The blue objects tend to have intrinsic brightnesses 
that are consistent with them being stars, rather than star clusters. The 
apparent absence of bright, compact, blue clusters is worth noting, 
as one might expect these to be present if, as suggested by Davidge (2008a), 
NGC 5102 experienced a burst of star formation during the 
recent past. Very few objects are detected with $r'-i' < 0$ and 
$i' < 24$ at large R$_{GC}$. This is a consequence of the drop in the number of bright 
MS stars in NGC 5102 at these radii, coupled with the modest number of background galaxies 
with this color at these magnitudes (e.g. Davidge 2008b).

	There is substantial contamination from background galaxies in the portion of 
the $(i', r'-i')$ CMD that contains RSGs, with the result that faint, unresolved 
galaxies introduce a significant source of uncertainty when probing the stellar content 
and structure of the low surface brightness outer regions of NGC 5102. A 
RSG plume, which is a conspicuous feature in the CMDs of spiral galaxy disks (e.g. Davidge 
2006; 2007), is not seen in the NGC 5102 CMDs. Still, the spatial distribution of objects 
with brightnesses and colors that are consistent with RSGs indicate that a modest 
population of these objects is present throughout the galaxy (\S 5). That bright MS stars 
and RSGs are found over a range of R$_{GC}$ indicates that there has been star formation 
throughout the NGC 5102 disk during the past few tens of Myr, although the small 
number of these objects indicates that the overall SFR during this time has been modest. 

\subsection{$(K, J-K)$ CMDs}

	The $(K, J-K)$ CMDs are shown in Figure 2. The WIRCam data 
sample the brightest AGB stars in NGC 5102, which have $K > 19$ and $J-K \approx 1.1$.
The density of stars on the infrared CMDs is lower than on the visible/red CMDs 
because evolution at infrared wavelengths is spread over a larger range of magnitudes, 
and only the most evolved AGB stars are detected in the WIRCam images (\S 3.4). 
Many of the brightest stars in the central 1 kpc are recovered in the WIRCam data 
because of the comparatively (at least with repect to visible wavelengths) high 
contrast between these objects and the underlying body of fainter stars. The diminished 
density of stars on the infrared CMD notwithstanding, an obvious concentration of 
AGB stars can be traced out to R$_{GC} \approx 4$ kpc in Figure 2. 

	The CMDs of sources with R$_{GC} > 4$ kpc are dominated by objects that do not 
belong to NGC 5102. The plume of objects with $K < 18$ and 
$J-K \approx 0.7$ consists of foreground Galactic disk stars, 
while the more diffuse sequence with $K > 16$ and $J-K \approx 1.6$ is made up of background 
galaxies. The WIRCam observations of NGC 5102 are spatially 
complete out to R$_{GC} \approx 6$ kpc, and the decrease in the 
number density of objects in the CMDs with R$_{GC} > 6$ kpc reflects the 
progressively smaller areas on the sky that are sampled as one moves to larger R$_{GC}$. 

\subsection{Comparisons with Isochrones}

	Isochrones from Girardi et al (2002; 2004) are compared with the CMDs 
of objects with R$_{GC}$ between 2 and 3 kpc in Figure 3. Davidge (2008a) found that 
the majority of RGB stars in the disk of NGC 5102 have 
[M/H] between --0.9 and --0.1, and so only models with Z = 0.019 and 
Z = 0.004 are considered here. While Kraft et al. (2005) deduce a super-solar metallicity 
using spectral synthesis techniques, this is based on observations of the central 
regions of NGC 5102, and not the disk (Davidge 2008a).

	The red stars sampled with these data are intrinsically bright and 
are highly evolved. Therefore, some aspects of the physics 
used to construct the portions of the isochrones 
that deal with the advanced stages of AGB evolution are worth noting. 
The isochrones used here include evolution during the thermally-pulsing AGB (TP-AGB) phase. This 
evolution is tracked using the models discussed in the appendix of Marigo \& Girardi 
(2001), which include prescriptions for hot-bottom burning and the third dredge-up.
There are uncertainties in key aspects of the model physics, such as the 
equation of state, the rates of mass loss among highly evolved stars, and the handling of
convection and convective boundaries. Girardi et al. (2000) discuss the model physics 
and the related assumptions for the portions of isochrones used here that cover evolution prior 
to the TP-AGB. The isochrones were transformed onto the observational plane by assuming 
oxygen-rich photospheres, and so do not match the properties of C stars. 
Finally, an observational complication of using highly evolved AGB stars as 
chronometers is that a large fraction are long period variables 
(LPVs). With photometric amplitudes that approach a magnitude in $K$, the presence of 
LPVs blur the boundaries of the region on the CMDs that contain the most highly evolved 
stars. 

	The MS turn-off is consistent with 
log(t$_{yr}) \leq 7.5$, indicating that star formation occured 
during the past few tens of Myr at intermediate radii in NGC 5102. 
As for more evolved stars, the age-metallicity degeneracy confounds 
efforts to determine ages for these objects from photometric data alone, and 
the ages estimated for AGB stars from the Z=0.004 models are 0.2 - 0.3 
dex older than those from the Z=0.019 models.

	C stars are important probes of stellar content because 
they have conspicuous photometric properties, and contribute 
significantly to the light output from simple stellar systems with ages between roughly 
1 Gyr (Persson et al. 1983) and $\approx 3$ Gyr (Cole \& Weinberg 2002; Feast et al. 2006). 
Demers \& Battinelli (2005) investigated the C star content in the outer regions 
of M31, with the majority of the candidate C stars identified 
on the basis of their $g'r'i'$ spectral-energy distributions. 
The area of the $(i', r'-i')$ CMD that Demers \& Battinelli (2005) found to contain C stars 
is indicated in the upper panels of Figure 3. The area containing oxygen-rich and 
carbon-rich AGB stars overlaps on the $(i', r'-i')$ CMD, and so C stars can not be identified 
from $i'$ and $r'$ observations alone. Still, it is evident from Figure 3 that some C stars may 
be lurking in the $(i, r'-i')$ CMDs of NGC 5102, and this is confirmed 
by the detection of C stars in the WIRCam observations (see below), 
where all but the warmest C stars have photometric properties that are very 
different from those of oxygen-rich giants.

	Line blanketing is much less of an issue at infrared wavelengths than at visible 
wavelengths, and one consequence is that the infrared CMDs sample 
stars that tend to be older than those in visible/red CMDs. 
Comparisons with the isochrones suggest that the majority of stars in the 
$(K, J-K)$ CMDs have log(t$_{yr}) \leq 9.0$. Still, there are stars 
near the red edge of the isochrones with $M_K \geq -8$ that {\it may} have ages approaching 
10 Gyr, although at least some of these are probably warm C stars (see below).

	The red envelope of the isochrones in Figure 3 occurs near $J-K = 1.25$ (Z = 0.004) 
and $J-K  = 1.3$ (Z=0.019), due to the assumption of oxygen-dominated photospheres. 
In fact, there are objects in the $(K, J-K)$ CMDs with $J-K \geq 1.3$, 
and some of these are probably C stars. The LMC $(K, J-K)$ CMD 
constructed by Nikolaev \& Weinberg (2000) from 2MASS data contains a prominent 
C star sequence, and the region of the LMC CMD that contains C stars is indicated 
in the lower row of Figure 3. Of course, some of the red objects in the NGC 5102 CMDs
are undoubtedly compact background galaxies, the majority of which will have $J-K > 1.0$. 
Still, in \S 5 it is shown that the spatial distribution of the objects with $J-K \geq 1.3$ 
cluster around the main body of NGC 5102, as expected if a large fraction are C stars.

\section{M Giants and C Stars as Probes of the Recent Star-Forming History of NGC 5102}

\subsection{A Constant SFR Model}

	The numbers of C stars and bright M giants in NGC 5102 can be used to gain insight 
into its evolution during the past few Gyr. For a given 
star-forming history, the ratio of C stars to M giants on the TP-AGB 
can be calculated using the fuel consumption theorum (e.g. Renzini \& Buzzoni 1986), 
and these predictions can be compared with the observed ratio. For the current work a model 
in which the SFR in the disk of NGC 5102 is constant with time, with a disk age of 
10 Gyr, is considered. There is evidence that the SFR in NGC 5102 has departed 
significantly from the time-averaged mean SFR during the past $\approx 1$ Gyr (e.g. Davidge 
2008a; Kraft et al. 2005), but the goal of the present effort 
is not to reproduce the observed statistics in NGC 5102. Rather, a constant SFR model 
simply provides a convenient benchmark for gauging the relative behaviour of the SFR with 
respect to time when comparisons are made by with observations.

	The relative contributions that stars evolving on the 
TP-AGB make to the total luminosity of a stellar system 
were computed using evolutionary fluxes and fuel consumption 
values from Maraston (2005). The fuel consumption fractions for 
C stars with Z$_{\odot}$/2 from Figure 12 of Maraston (2005), which was calibrated from 
C star counts in Magellanic Cloud star clusters, were also adopted. This has an important 
implication for C star statistics, since these fuel consumption values imply that all 
TP-AGB stars with ages between log(t$_{yr}$) = 9.0 and 9.3 are C stars (i.e. 
100\% of the fuel consumed during TP-AGB evolution is done so while the star has a C-rich 
atmosphere).

	Isochrones with ages log(t$_{yr}$) = 10 pass through the $(K, J-K)$ CMD 
(\S 3.4), and the numbers of old AGB stars in the data can be estimated using the 
fuel consumption theorum. Due to the progressive decline of the evolutionary 
flux with increasing age, coupled with the decline in the amount of fuel 
consumed by stars with M $\leq 2.5$ M$_{\odot}$ during TP-AGB evolution, then 
TP-AGB stars with ages in excess of 2 Gyr contribute only modestly ($< 10\%$) 
to the total number of oxygen-rich TP-AGB stars in the constant SFR model. 
Given the evidence for elevated levels of star formation during 
intermediate epochs then it is likely that the majority of 
bright AGB stars in the WIRCam data formed within the past $\approx 2$ Gyr.

	Using the assumptions described above then the total number of C stars and 
luminous M giants can be estimated if the total stellar mass of the NGC 5102 disk is 
known. The total brightness of NGC 5102 is $K = 6.9$ (Jarrett et 
al. 2003), so that M$_K = -20.6$. Assuming that the luminosity-weighted age of NGC 5102 
falls between 0.5 and 3 Gyr, which is the range of dates that Davidge (2008) and Kraft 
et al. (2005) suggest correspond to significant recent star formation activity, 
then M/L$_K \approx 0.2 - 0.5$ (Mouhcine \& Lancon 2003), and so the total stellar mass of 
NGC 5102 is $0.7 - 1.8 \times 10^9$ M$_{\odot}$. The surface brightness profile 
of Jarrett et al. (2003) suggests that roughly two-thirds 
of the $K$ light originates from the central portion of the galaxy that deviates 
from an exponential light profile, and so the disk mass is 
$2.5 - 6.2 \times 10^8$ M$_{\odot}$. Adopting the upper and 
lower luminosity limits for TP-AGB evolution from Mouhcine \& Lancon (2002) and a 
Salpeter IMF for the main sequence component, then the constant SFR model predicts 1800 -- 
4400 M giants experiencing TP-AGB evolution, and 4300 -- 10800 C stars.

	Given galaxy-to-galaxy variations in star-forming history and metallicity, one 
might not expect that the number of C stars predicted from the 
constant SFR model would agree with what is seen in other galaxies. 
It is thus interesting that the number of C stars predicted from the model, in which 
10\% of the total stellar mass formed between 1 and 2 Gyr in the past, is roughly 
consistent with C star counts in other galaxies after adjusting for differences in mass. For 
example, the LMC, which has a total stellar mass $\approx 10^9$ M$_{\odot}$, contains 10$^4$ C 
stars (Nikolaev \& Weinberg 2000), so $\approx 2500 - 6200$ C stars would be predicted in a 
galaxy with the same star-forming history as the LMC, but a mass comparable to the disk of 
NGC 5102. Demers, Battinelli, \& Letarte (2003) find 413 C stars in NGC 3109, and the 
number of C stars predicted for a galaxy with the same M$_K$ as NGC 5102 from these data 
is $\approx 7400$. Finally, the number of C stars predicted from the Letarte et al. 
(2002) survey of NGC 6822 after scaling to the mass of NGC 5102 is $\approx 10^4$.

\subsection{Comparing the Predictions with Observations}

	C stars and M giants that are evolving on the TP-AGB can 
be identified based on their location on the $(K, J-K)$ CMD. 
For C stars, the number of objects throughout the disk of NGC 5102 
in the region indicated in Figure 3 has been counted. The blue-edge of the C star sequence 
in Figure 3 is $J-K = 1.5$. However, there is some uncertainty in defining the 
blue edge of the C star sequence, and so counts were also made for objects with a blue 
edge at $J-K = 1.3$ and having the same $K$ magnitude boundaries as the C star region 
in Figure 3. The final C star count is the mean of the counts with the 
two $J-K$ cut-offs.

	As for the region of the $(K, J-K)$ CMD that contains M giants 
on the TP-AGB, the Girardi et al. (2002) isochrones indicate that the onset of the TP-AGB 
phase of evolution occurs near M$_K \approx -7.0 \pm 0.5$ for intermediate-age stars, while 
the AGB-tip peaks at M$_{K} = -9.5$. The isochrones also 
predict that M giants on the AGB have $J-K$ between 0.9 and 1.3. Thus, the 
number of sources with M$_{K}$ between $-7 \pm 0.5$ and M$_{K} = -9.5$ and between 
$J-K = 0.9$ and 1.3 have been counted; the mean of the counts with the lower M$_{K}$ 
set at --7.5, --7.0, and --6.5 is adopted as the number of TP-AGB M giants in NGC 5102. 

	Applying the methodology described above, and 
correcting for contamination from background galaxies using 
observations of a control field along the minor axis of NGC 5102 (the number of 
contaminating objects is modest when compared with stars in NGC 5102) then 
there are $1260 \pm 410$ M giants and $550 \pm 180$ C stars in the disk of NGC 5102. The 
uncertainties are dominated by the dispersion in counts arising from 
the different photometric selection criteria described in the previous paragraphs.

	The numbers of M giants agree with the range predicted in \S 4.1, and 
broadening the color boundaries for M giants on the CMD by $\pm 0.1$ magnitude 
increases the number of these stars by 10 -- 20\%, 
further improving the agreement with the constant SFR predictions. 
In contrast, the number of C stars is significantly lower than predicted by the 
constant SFR model. Highly evolved AGB stars may be obscured by circumstellar dust, and this 
might decrease the numbers of C stars. However, it can be anticipated that this will 
be a factor only for the stars that are nearing the end of their TP-AGB evolution, and 
it is unlikely that correcting for such stars will boost the numbers of 
C stars substantially. Taken at their face value, the numbers of M giants and C stars 
in the $(K, J-K)$ CMD suggest that $\approx 10\%$ of the total stellar mass in the disk formed 
within the past 1 Gyr, but that $< 20\%$ of the total stellar mass formed within the 
past 2 Gyr. The mean SFR in NGC 5102 during the past 1 Gyr is then $\approx 1$ M$_{\odot}$ 
year$^{-1}$, whereas Davidge (2008b) estimated a SFR of at least 1.4 M$_{\odot}$ 
year$^{-1}$ based on modelling of the bolometric LF of AGB stars along the minor axis of 
NGC 5102. Given the (substantial) uncertainties in computing the mass of NGC 5102 from 
photometric data, the estimated SFRs are not significantly different.

	While there are considerable uncertainties in the computation of the 
numbers of M giants and C stars, such as determining the mass of the NGC 5102 disk, the 
ratio of C stars to M giants (C/M) should yield a more robust means of probing recent star 
formation. The counts from the CMDs indicate that C/M $= 0.4 \pm 0.2$, which is 
significantly smaller than the ratio predicted from the constant SFR model, which is C/M = 
2.4. The measured C/M ratio is consistent with a recent marked increase in the SFR of 
NGC 5102 during the past $\approx 1$ Gyr, which produced a high number of bright M 
giants when compared with the C stars that formed 1 -- 2 Gyr in the past. However, the 
reader is cautioned that a lower than expected C/M ratio could result from other factors. 
For example, a C/M ratio that is lower than the model
prediction could occur if the dominant population in NGC 5102 is either younger than 
$\approx 0.2$ Gyr or older than $\approx 2 - 3$ Gyr. Such ages for the dominant 
population are in conflict with the large number of intrinsically bright AGB stars 
that are seen throughout the disk of NGC 5102 (e.g. Davidge 2008a), and the evidence of elevated 
SFRs in other parts of the galaxy a few tenths of a Gyr in the past (e.g. Kraft et al. 2005).

	The C/M ratio also depends on metallicity. 
As discussed by Maraston (2005), the fraction of fuel that is consumed 
during TP-AGB evolution while a star has a C-rich atmosphere scales with 
metallicity, as a metal-rich star must dredge up more C in order to 
bind O in CO than a metal-poor star. Kraft et al. (2005) conclude that the 
central regions of NGC 5102 have a super-solar metallicity, and so a C star survey of this 
part of the galaxy should find a C/M ratio that is lower than in 
a comparatively metal-poor environment that had the same recent star-forming history. 
This being said, it is unlikely that metallicity plays a major role in the low 
C/M ratio in the disk of NGC 5102. The metallicity measured by Kraft et al. (2005) 
is based on the central regions of the galaxy, and the colors of the RGB sequence in the disk 
measured by Davidge (2008a) are indicative of a sub-solar metallicity. In fact, the peak 
metallicity among RGB stars is [M/H] $\approx -0.6$ (Davidge 2008a), 
which is lower than the metallicity used to compute the numbers of C stars and M giants 
in \S 4.1. The adoption of a lower metallicity in the calculations will result 
in a higher predicted C/M ratio, increasing the difference with respect to the observed ratio.
 
\section{THE SPATIAL DISTRIBUTION OF STARS}

\subsection{Mapping the Spatial Distributions of Stellar Types in NGC 5102}

	Stars with ages from $\approx 10$ Myr to at least a few Gyr are detected 
in the MegaCam and WIRCam images, and these data can be used to investigate 
the spatial distribution of star-forming activity with respect to time in NGC 5102. 
A caveat is that the ability to resolve stars that formed over a given age range 
drops as progressively older stars are considered, and 
isolating sites in a galaxy that formed stars over narrow age ranges becomes problematic.
Stars of different ages and evolutionary stages were identified
based on their locations in CMDs, as indicated in Figure 4. The 
bright MS stars and RSGs that formed during the past few hundred Myr are identified 
from the MegaCam data, while AGB stars that formed within the past few Gyr are identified 
from the WIRCam data. The region of the MegaCam CMD that contains RSGs has been 
divided into three groups, the brightest of which contains stars 
with ages $\leq 30$ Myr, whereas the faintest samples stars with ages in excess of 100 Myr.

	The various stellar types have distinctly different spatial distributions, and this 
is demonstrated in Figure 5. While there are only a modest number of objects in each group, 
it can be seen that MS stars and BRSGs occur in largest numbers in the south-western half 
of the disk, which is the area of the galaxy that also contains HII regions. In contrast,
the IRSGs and FRSG/BAGBs , where `BAGB' refers to `Bright AGB' stars, are more evenly 
distributed across NGC 5102. The uniform distribution of these stars is likely a consequence
of their ages. With ages that exceed many tens of Myr, these stars 
have probably obtained random velocities through interactions with molecular 
clouds that have allowed them to diffuse away from 
their places of birth, thereby blurring any structure in their original distribution. 

	The AGB stars with oxygen-rich (OAGB stars) and carbon-rich (CAGB stars) atmospheres 
that were identified from the WIRCam images define a more compact distribution than 
the MegaCam-based samples. Given that these stars have ages of a few hundred Myr to 
a few Gyr, then this suggests that star formation at this epoch was more centrally 
concentrated than during the past few hundred Myr. It is thus interesting that, in 
contrast to the spatial distribution of objects identified from the 
MegaCam data, the OAGB stars are not uniformly distributed. 
An arm of OAGB sources extends to the north east of the 
galaxy along the major axis, and this same feature is seen among 
CAGB stars. A corresponding feature is not seen to the south west of the galaxy. 

	There is a clear concentration of OAGB stars to the west of the galaxy center. 
Although the small numbers of CAGB stars makes the identification of 
an obvious concentration amongst these stars more difficult, the 
distribution of CAGB stars is clearly asymmetric along the east-west axis, in the sense 
that CAGB stars extend $\approx 2$ kpc to the west of the galaxy center, but 
only $\approx 1$ kpc to the east. A corresponding structure is not evident in the 
FRSG/BAGB counts as most of the area containing the OAGB concentration falls in a 
gap between CCDs. 

	The concentration of AGB stars coincides roughly with the 
center of a loop of [OIII] emission discovered by McMillan et al. (1994), for which they 
estimated an age of $\sim$ 10 Myr based on its size and expansion velocity. This might 
suggest that the AGB concentration was also a site of very recent star formation. 
However, while Davidge (2008) found MS stars to the west of the galaxy 
center, these tend to be fainter, and hence older, than the 
MS stars in the southern portions of NGC 5102.
In fact, the distribution of stars in Figure 5 indicates that the area near the AGB cluster 
does not contain an anomolous number of bright MS stars, with the caveat that 
the center of the AGB cluster falls in a gap between CCDs. The absence of HII regions in 
this part of NGC 5102, based on the entries in Table 4 of McMillan et al. (1994), 
further argues that it is not a site of star formation at the present day.

	The stellar content of the AGB concentration 
appears to differ from that of the NGC 5102 disk, and this is demonstrated in Figure 6. 
The $(K, J-K)$ CMDs of 1.6 arcmin$^2$ regions that are centered 50 arcsec 
immediately to the west (`AGB Cluster') and east (`Disk East') of the galaxy nucleus 
are compared in the upper row of Figure 6. There are almost twice as many stars in the 
AGB Cluster area, and so the CMD of this feature 
is more richly populated than that of the eastern disk. 
The CMD of the Disk East field also appears to be broader than 
that of the AGB Cluster region. Indeed, the Disk East field contains 6 stars with 
$J-K > 2$, whereas there is only one such star in the AGB Cluster CMD. 
The more compact CMD morphology of the AGB Cluster is consistent 
with this area being dominated by stars that have a narrower range in age and metallicity 
than occurs in the main body of the disk.

	The $J-K$ color distributions of stars with $K$ between 19 and 20 (i.e. with M$_K$ 
between --7.5 and --8.5) in the Disk East and AGB Cluster fields are compared in the lower 
row of Figure 6. Both distributions peak near J--K = 1.1. However, the color distribution of 
the AGB Cluster appears to be skewed by $\approx 0.2$ mag 
to smaller J--K colors than that of the Disk East field. 
A Kolomogorov-Smirnov test indicates that the color distributions differ 
at the 91\% significance level, and so the evidence for a difference in stellar content 
should be considered to be suggestive, rather than conclusive. Deeper photometric studies 
will provide greater statistical robustness when comparing the stellar content 
of the AGB Cluster with that of the rest of the disk. This being said, if 
the stars in both fields have similar metallicities, then a 0.2 mag difference 
in J--K is consistent with a younger mean age for the AGB Cluster field, 
amounting to $\Delta$log(t$_{year}) \geq 0.5$ dex, based on the isochrones shown in 
Figure 3. 

\subsection{The Distribution of Stars Along the Major Axis}

	The distribution of stars in the outermost regions of the NGC 5102 disk may provide 
clues into its past history. Because the disk of NGC 5102 is viewed at a moderately high 
angle, the projected intrinsic width of the disk complicates efforts to study 
the distribution of stars, especially at small R$_{GC}$ along the minor axis. To 
reduce the possible impact of disk thickness on structural studies, in the following 
analysis star counts are restricted to a strip along the major axis with a projected 
width $\pm 1$ kpc. 

	The numbers of MS, BRSG, and IRSG sources are too small to allow their radial 
profiles to be investigated. However, FRSG/BAGB, OAGB, and CAGB stars are present 
in moderately large numbers, allowing limited conclusions to be drawn about their
radial distributions. The number of sources in these groups 
are shown as a function of R$_{GC}$ in Figure 7. The counts have 
been corrected statistically for contamination from background galaxies and foreground stars 
by subtracting number counts measured in control fields. 
The CMDs in Figure 1 indicate that incompleteness is an issue in 
the FRSG/BAGB star counts for R$_{GC} < 3$ kpc, and so little weight should be 
assigned to the FRSG/BAGB counts near the center of NGC 5102.

	It is evident from Figure 7 that the OAGB and FRSG/BAGB stars have very 
different radial distributions, in that the former are more 
centrally concentrated than the latter. This difference 
is perhaps not surprising given that there is not a one-to-one correspondence between the 
stars that fall in the FRSG/BAGB portion of the $(i', r'-i')$ CMD and those located 
in the OAGB portion of the $(K, J-K)$ CMD, and that stars with different age 
distributions are found in each of these boxes. As noted in \S 5.1, the more compact 
spatial distribution of OAGB stars suggests that star formation $\approx 1$ Gyr ago was more 
centrally concentrated than during the past $\approx 0.1$ Gyr.

	The number density of FRSG/BAGB stars in Figure 7 flattens when R$_{GC} > 5$ kpc, 
although (1) there are substantial errors in the measurements 
at large R$_{GC}$ due to the small numbers of objects, (2) 
even a modest variation in the number of background galaxies will have a major 
impact on the behaviour of the FRSG/BAGB profile at large R$_{GC}$, and (3) the distribution 
of stars at large radii is asymmetric in both the WIRCam (\S 5.1) and MegaCam (\S 5.3) 
datasets. The radial profile of CAGB stars, which have systematically different 
photometric properties than the majority of FRSG/BAGB objects, also show evidence of 
flattening near R$_{GC} \approx 5$ kpc, although the WIRCam data are limited 
in their spatial coverage on the sky and so star counts can not 
be extended past this radius. Still, the number density of 
OAGB stars drops off near R$_{GC} \approx 4.5$ kpc, and the OAGB and CAGB profiles in Figure 7 
suggest that the ratio of C stars to M giants increases with R$_{GC}$ in NGC 5102, 
as is seen in other disk galaxies (e.g. Demers \& Battinelli 2005).

	The FRSG/BAGB and CAGB profiles in Figure 7 suggest that the distribution of stars 
in the disk of NGC 5102 does not follow a simple exponential profile. 
Erwin, Beckman, \& Pohlen (2005) investigate the light profiles of barred 
early-type galaxies, and find that at least 25\% of their sample have 'anti-truncated' 
light profiles. The light profiles of these galaxies 
at large radii follow an exponential profile that is shallower than at small radii. 
Similar light profiles are seen among unbarred disk galaxies, with the highest frequency 
of occurence among galaxies with early morphological types (Pohlen \& Trujillo 2006). 

	Two possible mechanisms for generating an anti-truncated light profile are 
considered here, with the goal of yielding possible insights into the past history of 
NGC 5102. Anti-truncated light profiles may be the consequence of mass 
re-distribution in disks during interactions. Younger et al. (2007) argue that the angular 
momentum of gas that is funneled to smaller radii is transfered to disk stars, causing 
the latter to move outwards. To the extent that the disks of `normal' spiral galaxies 
likely have formed stars over a wide range of ages prior to any interactions, 
then most of the stars that move outwards in a spiral galaxy will have 
ages in excess of a Gyr. If the outer disk of NGC 5102 was populated in this way 
then it should have a luminosity-weighted age that is older than that of the inner disk, 
which at some point (or points) in the past experienced an elevated SFR due to the 
interaction. 

	The model described by Younger et al. (2007) is not without its problems. The 
distribution of stars in the outer disk of the interacting galaxy M81 does not support 
the predicted stellar content trends, in that the outer 
disk of M81 appears to be {\it younger} than the inner disk. 
This might indicate that some fraction of the gas in M81 was the beneficiary of angular 
momentum re-distribution, or that gas was accreted from M82 (Davidge 2009). In addition, 
secular processes, such as viscosity-induced re-distribution of angular momentum, are 
expected to act on the order of a few disk crossing times, and these will blur any age 
signatures that may have been imprinted at the time of formation of the outer disk.

	An anti-truncated light profile may also result from the accretion of 
material with an angular momentum distribution that is different from that of the 
original disk, and this provides a natural explanation for warped outer disks 
(van der Kruit 2007). If this mechanism is/was at play in NGC 5102 then the 
oldest stars in the outer disk will be younger than those in the main body of the disk. 
A difference in age between the inner and outer disk of NGC 5102
might be measureable using integrated light or the brightest resolved stars if the 
outer disk was assembled from material that was accreted only during the past 1 -- 2 Gyr. 
If the outer disk was accreted earlier then very deep imaging 
will be required to detect an age difference between the inner and outer disk. 
Secular processes will also blur differences in stellar content with time.

	We close this section by noting that an extended disk 
might also result from star burst activity, although it is 
not clear if an anti-truncated light profile would result. Dalla Vecchia \& Schaye (2008) 
model galaxy outflows, and find that the gas dispelled by winds initially escapes the 
disk plane along the minor axis. However, material builds up along the minor axis as gas 
loss proceeds, with some gas falling back into the disk plane. The 
resulting increase in density forces gas to escape along the major axis, and the disk 
boundary moves outwards as ejected gas moves to larger radii. A prediction of this model 
is that the mass-weighted age of the region near the outer 
boundary of the disk after the star burst subsides will be younger 
than at larger radii, with the oldest stars in this part of the galaxy having 
an age that corresponds to the onset of material being ejected along the major axis. In the 
case of NGC 5102 this means that the outer disk will be populated by stars with ages 
$< 1$ Gyr. There should then be an extended distribution of OAGB stars, which is not seen.

\subsection{The Stellar Distribution at Large Radii}

	Regions of a galaxy that have very low stellar density and span large spatial scales 
can be identified from star counts if the number density of photometrically-selected objects 
exceeds a threshold defined by the number of foreground stars and background galaxies. 
Davidge (2008b) searched for young stellar groupings in the M81 -- M82 debris field, and 
found three objects that have surface brightnesses of 27 -- 28 mag arcsec$^{-2}$ in $V$. 
This procedure has been applied to the MS and FRSG/BAGB objects in the MegaCam data. Source 
counts were made in $700 \times 700$ pixel bins, which corresponds to $130 \times 130$ 
arcsec, or $2.1 \times 2.1$ kpc at the distance of NGC 5102. This binning produces a 
moderately large number of objects per resolution element, while also sampling the spatial 
scales over which significant changes in galaxy structure might be expected. 
The results are not critically sensitive to bin 
size, and the basic conclusions drawn below do not change if $500 \times 500$ 
pixel or $1000 \times 1000$ pixel binning factors are employed.

	Aside from the central few arcmin of NGC 5102, statistically significant 
over-densities were not found in the numbers of MS objects 
over the 1 degree$^2$ MegaCam field. This is interesting given the low level 
of contamination from background galaxies with blue colors (Davidge 2008b). 
This null detection suggests that NGC 5102 is not accompanied by kpc-sized companions 
that contain stars that formed within the past few tens of Myr.

	The situation is very different when the spatial distribution of FRSG/BAGB objects 
is considered. A gray scale image illustrating the distribution of FRSG/BAGBs is shown in 
the top panel of Figure 8. The star counts in the north west portion of the image are 
affected by bright stars, which suppress the detection of faint sources over arcmin or 
larger scales. The brightest of the stars in the MegaCam image is the $V = 2.8$ star HD 
112892 ($\iota$ Cen), which is at the center of the large dark spot in Figure 8.
Still, three large-scale features can be seen. The most conspicuous of these, at the center 
of the map, is the main body of NGC 5102, where the density of FRSG/BAGB objects 
is highest. The second feature falls along the south west arm of the major axis, 
where elevated source counts can be traced out to $\approx 15$ kpc ($\approx 16.6$ scale lengths) 
at the distance of NGC 5102. The third structure is a concentration 
of FRSG/BAGBs along the south east arm of the minor axis, that extends out to a projected 
distance of $\approx 18$ kpc. 

	What is the statistical significance of the features identified in Figure 8? 
To answer this question, the standard deviation about the mean number of counts 
pixel$^{-1}$ was measured in the southern quarter of the FRSG/BAGB map. 
This region was adopted as the control area since contamination 
from stars associated with NGC 5102 should be low. The pixels where the number of FRSG/BAGBs 
exceeds the mean in the control field at the 2, 3, 4, and 5$\sigma$ levels 
are shown in the bottom row of Figure 8. 

	Not unexpectedly, the FRSG/BAGB counts in the inner regions of NGC 5102 have 
the highest statistical significance, exceeding those in the control field at more than 
the $5\sigma$ level. What is of greater interest for the outer regions of NGC 5102 
is that the FRSG/BAGB number counts in pixels along the south west major axis 
exceed those in the control area at more than the $3\sigma$ level. 
This is in direct contrast to the number counts along the north 
west arm of the major axis, which do not exceed those in the control area 
at even the $2\sigma$ level. The distribution of stars in the outer disk of NGC 5102 is thus 
not homogeneous. Rather, there is clumpy sub-structure, such that the stellar densities 
along the south west arm of the major axis in the outer disk are significantly 
higher than those along the north east arm. Finally, pixels along the south east 
arm of the minor axis exceed those in the control area between the 2 and $3\sigma$ 
levels; while having lower statistical significance than the structure on the south west 
major axis, these pixels are clustered together, suggesting that their elevated counts are 
not statistical flukes.

	The surface brightnesses of the areas along the major and 
minor axes of NGC 5102 can be estimated if their 
stellar content is assumed to be the same as areas of the NGC 5102 disk where 
the surface brightness has been measured directly. The assumption of similar stellar content 
is certainly open to debate given the evidence for population gradients in disks 
(e.g. Bell \& de Jong 2000). Still, this issue can be 
mitigated to some extent by using fields at moderately large R$_{GC}$ in NGC 5102 to 
calibrate surface brightness.

	A $2\sigma$ detection in the number counts corresponds 
to $7 \times 10^{-4}$ FRSG/BAGB objects arcsec$^{-2}$, whereas a $3\sigma$ detection 
corresponds to $1 \times 10^{-3}$ FRSG/BAGB stars arcsec$^{-2}$. From Figure 7 
it is seen that at R$_{GC} = 7.5$ kpc, the mean density of FRSG/BAGB stars is $\approx 1.1$ 
arcmin$^{-2}$, or $3.1 \times 10^{-3}$ arcsec$^{-2}$. The surface brightness at this 
radius is $\approx 25.5$ mag arcsec$^{-2}$ in $B$ (Figure 2 of van 
Woerden et al. 1993). Therefore, features that exceed the FRSG/BAGB number counts 
in the control field at the $3\sigma$ level have a surface brightness 
$\approx 26.7$ mag arcsec$^{-2}$, whereas those that exceed the control field counts at the 
$2\sigma$ level have a surface brightness of 27.1 mag arcsec$^{-2}$. These 
very low surface brightnesses explain why these objects have eluded detection until now.

\section{DISCUSSION \& SUMMARY}

	The bright stellar content of the nearby (D $\approx 3.2$ Mpc) S0 galaxy NGC 5102 has 
been surveyed with the WIRCam and MegaCam imagers on the CFHT. 
With the exception of the crowded central regions of the galaxy, the MegaCam data 
cover all of NGC 5102, while the WIRCam data are spatially complete out to R$_{GC} \approx 
6$ kpc. This is the first published study of the stellar content of NGC 5102 at 
near-infrared wavelengths, and the data have been used to probe the spatial distribution 
of stars, with particular emphasis on the outer regions of the galaxy. Structural 
information of this nature will provide insights into the past evolution of NGC 5102, and 
may identify the trigger of the elevated star-forming activity that occured within the 
past 1 Gyr. 

	There is a large body of literature that deals with the properties of gas-poor 
disks, and most of the papers focus on the higher mass kindred of NGC 5102. Given that the 
current paper deals with only a single galaxy, no attempt is made to present a 
comprehensive review of the literature on S0 galaxies. Rather, we simply note here that
it is broadly accepted that the progenitors of S0 galaxies are spiral galaxies that have 
experienced gas depletion. There is an environmental dependence (e.g. Butcher \& Oemler 
1984), and the mechanism causing gas removal has been the subject of much debate, focusing 
on galaxy-galaxy interactions (e.g. Lavery \& Henry 1988), or infall from the field into an 
area with a dense intergalactic medium (e.g. van Dokkum et al. 1998). An 
active nucleus may also drive gas from a galaxy (e.g. Silk 2005; Silk \& Norman 2009), 
although this is expected to be restricted to disk systems that have massive spheroids. Of 
course, it is possible that not all S0s were subjected to the same gas-removal mechanism.

	Low surface brightness structures such as those found here in the vicinity of NGC 
5102 may prove challenging to detect in more distant S0 galaxies, and so NGC 5102 may play 
an important role for investigating the formation and evolution of gas-depleted disks. This 
being said, the reader is reminded that some characteristics of NGC 5102 are not typical 
of S0 galaxies. These include its intrinsic faintness (M$_B \approx -17.7$; Karachentsev et al. 
2002), the blue central colors and deep Balmer absorption lines (Gallagher et al. 1975), the 
modest total dust mass (van Woerden et al. 1993), the low number of x-ray binaries, and the 
low globular cluster specific frequency (Kraft et al. 2005). In regards to 
the modest intrinsic brightness and blue central colors 
of NGC 5102, Bedregal et al. (2008) investigate the spectroscopic 
properties of Fornax cluster S0 galaxies that span a range of masses, and find a trend 
between luminosity-weighted central age and galaxy mass, in the sense that lower mass S0's 
have younger central ages. The blue central colors of NGC 5102 may thus be tied 
to its modest mass.

\subsection{Tracing Star Formation in NGC 5102: The Recent Past and the Immediate Future}

	There is only a modest number of resolved bright MS stars and RSGs throughout the 
disk of NGC 5102, demonstrating the low SFR during the past $\approx 30$ Myr. The majority of 
young stars are in the southern half of the galaxy, which is where the only HII regions 
are located. This is also where the HI density is highest and there are HI clumps; 
for comparison, HI in the northern part of the NGC 5102 disk is more 
uniformly distributed, with a lower mean density (van Woerden et al. 1993). 
The HI distribution strongly suggests that -- barring some external influence -- the 
part of the disk that is presently south of the galaxy center will continue to be the 
dominant area of star formation for time scales of at least a few disk crossing times.

	RSGs and AGB stars span a much wider range of ages than the brightest 
MS stars, and these objects are more uniformly distributed throughout the disk
of NGC 5102 than bright MS stars. This indicates that star formation 
within the past $\approx 0.1 - 0.2$ Gyr was not restricted to selected locations in the disk. 
Indeed, the number densities of FRSG/BAGB stars at intermediate radii along the 
northern and southern arms of the major axis are similar, indicating that the northern 
and southern parts of the galaxy had similar time-averaged SFRs $\approx 0.1 - 0.2$ Gyr 
in the past. 

\subsection{The Outer Disk of NGC 5102}

	The number counts of objects that are classified as 
FRSG/BAGB stars suggest that the light profile of the NGC 5102 disk does not follow a 
single power law, but breaks to a shallower profile near 
R$_{GC} \approx 5 - 6$ kpc (5.5 -- 6.7 disk scale lengths). 
The error bars in the FRSG/BAGB counts are significant at large radii, and 
background galaxies that -- at an angular resolution 
of roughly an arcsec -- are unresolved introduce a considerable source of 
uncertainty. Still, the majority of early-type disk galaxies have 
light profiles that become shallower at large radii (Pohlen \& Trujillo 2006). The spatial 
distribution of OAGB stars, selected from the WIRCam data, is more compact than that of 
the FRSG/BAGB stars. Given that the OAGB stars tend to be older than those in the FRSG/BAGB 
sample, then this suggests that the outer disk of NGC 5102 has a younger mean age 
than the inner disk. This is similar to what is seen in M81, where the structural 
properties of the outer disk have almost certainly been affected by 
interactions with M82 (Davidge 2009).

	Stars along the major axis of NGC 5102, which presumably belong to the disk, can be 
traced out to R$_{GC} \approx 14$ kpc, or $\approx 15.5$ disk scale lengths, in the southern half 
of the galaxy. This indicates that the disk may extend to a greater distance than was 
estimated by Davidge (2008a), who used RGB star counts along the minor axis to trace 
the disk out to a de-projected radius of $\approx 10$ kpc. 
Clumpiness in the outer disk of NGC 5102 may explain why the disk 
radius estimated made by Davidge (2008a) along the minor axis differs from that 
found here. The distribution of FRSG/BAGB stars in Figure 8 indicates that there is 
structure in the outer disk of NGC 5102, with a marked concentration of objects along the 
south west semi-major axis (\S 4.3). The disk of NGC 5102 is also 
markedly warped at large radii in the northern portion of 
the galaxy. This is evident in the HI map of van Woerden et al. (1993), where the HI 
distribution veers to the north west of the northern arm of the major axis 
as extrapolated from smaller R$_{GC}$. OAGB stars follow the HI warp.
A cluster of background galaxies could masquerade as the southern extension of 
the NGC 5102 disk, although such a cluster would have to be positioned 
fortuitously along the major axis of NGC 5102.

	Deep images with angular resolutions of the order 0.1 arcsec or better will provide 
the data needed to confirm the nature of the objects that define the 
low surface brightness structures along the major axis of NGC 5102. 
If, as suggested here, the structures are made up of stars that belong to the disk of 
NGC 5102, as opposed to background galaxies, then the projected density of extended objects 
in these structures will be comparable to that in other parts of the area around NGC 5102. 
On the other hand, if high angular resolution observations detect an excess population of 
galaxies then the FRSG/BAGB counts will need to be re-visited.
 
\subsection{The Nature of the AGB Concentration Near the Center of NGC 5102: Star Cluster or Satellite?}

	The kpc-sized concentration of AGB stars immediately to the west of the galaxy 
center is the most prominent structure found in the vicinity of NGC 5102, and the nature of 
this feature is of considerable importance for understanding the recent evolution of NGC 
5102. The relatively blue $J-K$ colors of the majority of AGB stars in this area (\S 5) 
suggest that this was probably an area of localized star formation in the not-to-distant 
past. The physical coincidence with a probable superbubble (McMillan et al. 1994) suggests 
that star formation may have continued in this stellar concentration up to within the past 
few tens of Myr, and HI is concentrated in this area (van Woerden et al. 1993). Deeper 
images should reveal a population of intermediate age MS stars in the AGB concentration 
with a spatial density that is roughly twice that on the eastern side of the disk. The 
age measured from the main sequence turn off of this stellar concentration will provide a 
means of timing star burst activity in NGC 5102.

	Two possible origins for the AGB concentration are considered here. 
One is that the AGB concentration is a large cluster in the disk of NGC 5102. 
An awkward aspect of this interpretation is that much of 
the most recent star-forming activity during intermediate epochs in NGC 5102 would have been 
concentrated in a single region of the disk that is offset from the center of the galaxy. 
This is not typical for starburst systems, where star formation tends to be centered on the 
nucleus (e.g. Kewley, Geller, \& Barton 2006, and references therein). 
Still, if the star cluster interpretation is correct then it can not be older 
than $\approx 1$ Gyr. The crossing time of the NGC 5102 disk is on the order of a few 
hundred Myr, and coherent structures might be expected to dissipate after a few crossing 
times, due to kinematic heating by interactions with large gas clouds and other star 
clusters. The difference in colors between the brightest M giants in the disk to the east of 
the galaxy center and the AGB concentration (Figure 6) is qualitatively 
consistent with a relatively `young' age for the AGB concentration, 
although it is difficult to assign a reliable absolute age from these data alone. 

	The second interpretation is that the stellar concentration 
is a satellite galaxy seen in projection against (or behind) the disk. 
Kraft et al. (2005) hypothesize that a gas-rich dwarf galaxy 
triggered the large episode of star formation in NGC 5102 during intermediate epochs. 
A long-recognized difficulty with this hypothesis is that NGC 5102 is in an isolated 
environment. Using distances measured by Karachentsev et al. (2007), the nearest known 
neighbor of NGC 5102 is the barred spiral galaxy ESO383-G087, which is at a distance 
of $0.3 - 0.4$ Mpc. The satellite interpretation of the AGB concentration is thus attractive 
as it provides a trigger for the burst of star formation in NGC 5102. 

	Comparatively easy tests of the satellite hypothesis are that (1) a satellite 
would have left a trail of material as it was disrupted, and this trail might still be 
visible, and (2) NGC 5102 should show signs of having experienced a recent interaction. 
While ambiguous stellar streams have not been detected near NGC 5102, there are two 
{\it possible} tidal remnants. The grouping of stars along the minor axis could 
be the remnant of an interaction involving a satellite with an orbit that was perpendicular 
to the disk plane. The integrated brightness of the three minor axis pixels 
in Figure 8 that breech the $2\sigma$ threshold is V $\approx 15.3$, such that M$_V \approx 
-12.2$, which is $\approx 1\%$ of the integrated brightness of NGC 5102. Mori \& Rich (2007) 
model interactions in which satellites with masses that are no more than a few percent of 
the total mass of M31 pass through the central regions of M31. While caution should of 
course be exercised when extending these results to NGC 5102, some broad conclusions are 
worth noting. The satellite is significantly disrupted, and its stars are distributed 
throughout the extraplanar environment, forming distinct streams and shells 
during the first $\approx 1$ Gyr. The disk of M31 is not affected 
significantly as long as the satellite has a mass that is 
less than $\approx 1\%$ of that of the galaxy. A few Gyr after the first interaction the 
extraplanar region is dominated by a diffuse component, with little evidence of streams 
and shells.

	The clumpy structures found in the outer disk of NGC 5102 are reminiscent 
of the extended outer disk of M31 (Ibata et al. 2005). Based on its disk-like 
stellar content, Faria et al. (2007) argue that the M31 `cluster' G1, which falls along 
the major axis of M31, might actually be a fragment of the M31 disk that was pulled from the 
disk during a past interaction. Richardson et al. (2008) find that the stellar 
content throughout the outer disk of M31 is highly uniform, and argue that the 
extended disk may be the consequence of the heating and 
disruption of the thin disk of M31 by a satellite. An extended disk formed in 
this manner may stay in place for many Gyr given the long mixing times in the outer 
regions of galaxies (e.g. Johnston, Hernquist, \& Bolte 1996). It is further worth 
noting that the ring-like HI distribution in NGC 5102 is reminiscent of the 
distribution of gas and dust in M31, which simulations suggest may be the result 
of a smaller companion (presumably M32) passing through the disk (e.g. Block et al. 
2006). The warping of the NGC 5102 disk, which is seen in both the HI and stellar 
distributions, may also be the consequence of a tidal encounter. 

	There are other tests of the satellite hypothesis, but they will prove challenging 
to implement. A satellite galaxy would probably contain a substantial 
old population, with a metallicity that may be distinct from that of NGC 5102. 
However, such a population would prove difficult to detect given that the proposed 
satellite is viewed against/through the crowded inner regions of NGC 5102. Another test 
is that the radial velocities of stars in the cluster and the disk may differ. An obvious 
difficulty is that an investigation of the velocities of even the most luminous stars 
will require larger telescopes than those currently available. This being said, 
the AGB concentration is coincident with a localized peak in the HI distribution, and this 
feature is not kinematically distinct in HI channel maps (van Woerden et al. 1993). 
A large velocity difference between NGC 5102 and the supposed satellite may not be present 
today, as dynamical friction will act to harmonize the kinematic properties of stars 
in the satellite with those in NGC 5102, although this may require very long time scales 
depending on the orbital geometry and relative system masses (e.g. Colpi, Mayer, \& 
Governato 1999).

\subsection{Star Formation in an Outflow from NGC 5102?}

	In \S 6.3 the collection of stars at a projected galactocentric distance of 
$\approx 18$ kpc on the south east minor axis of NGC 5102 was discussed in the context of 
it being a possible tidal remnant. In this section we consider another possible origin for 
this structure. Galaxies with elevated SFRs may experience outflows, and this 
can have an impact on their surroundings. The outflow from M82 may 
trigger star formation in its extraplanar regions (e.g. Davidge 2008c,d), 
as the wind interacts with circumgalactic clouds that 
presumably were tugged from M82 and/or M81 as they interacted, or condensed from material 
that was ejected from M82 by winds. The features identified as 
the `Cap' (Devine \& Bally 1999) and M82 South (Davidge 2008d) are perhaps the most obvious 
signatures of the interplay between the M82 outflow and surrounding material.

	The ISM of NGC 5102 shows only marginal evidence for an outflow at 
the present day (Shwartz et al. 2006), although Kraft et al. (2005) find 
diffuse X-ray emission in the central 1 kpc of NGC 5102, which they 
attribute to a superbubble that is powered by SNe associated with the most recent burst of 
star-forming activity. However, the SFR in NGC 5102 during intermediate epochs was probably 
sufficient to power a galaxy wind (\S 7.3 of Davidge 2008a). Could the diffuse stellar 
component along the south east minor axis of NGC 5102 have formed in such an outflow?

	An outflow origin for these stars provides a natural explanation 
for their location along the minor axis of NGC 5102. However, such structures will 
probably not be long-lived, and will probably be disrupted over a few orbital timescales 
around NGC 5102 (i.e. within $\approx 1$ Gyr). There would also have to be a large diffuse 
gas component surrounding NGC 5102, and no evidence for this has been found. 
Finally, with M$_V \approx -12.3$ (\S 6.3) the minor axis structure 
would be much larger than the largest outflow cluster in M82 
(Davidge 2008d). Thus, while an outflow origin for these stars can not be ruled out, 
this possibility appears to be unlikely.

\acknowledgements{It is a pleasure to thank the Director of the CFHT, Christian Veillett, 
for allocating the discretionary time during which these data were recorded. Thanks are 
also extended to Sidney van den Bergh for commenting on an earlier version of the 
manuscript, and to the anonymous referee, whose comments led to significant improvements in 
the paper.}

\clearpage
\begin{figure}
\figurenum{1}
\epsscale{0.75}
\plotone{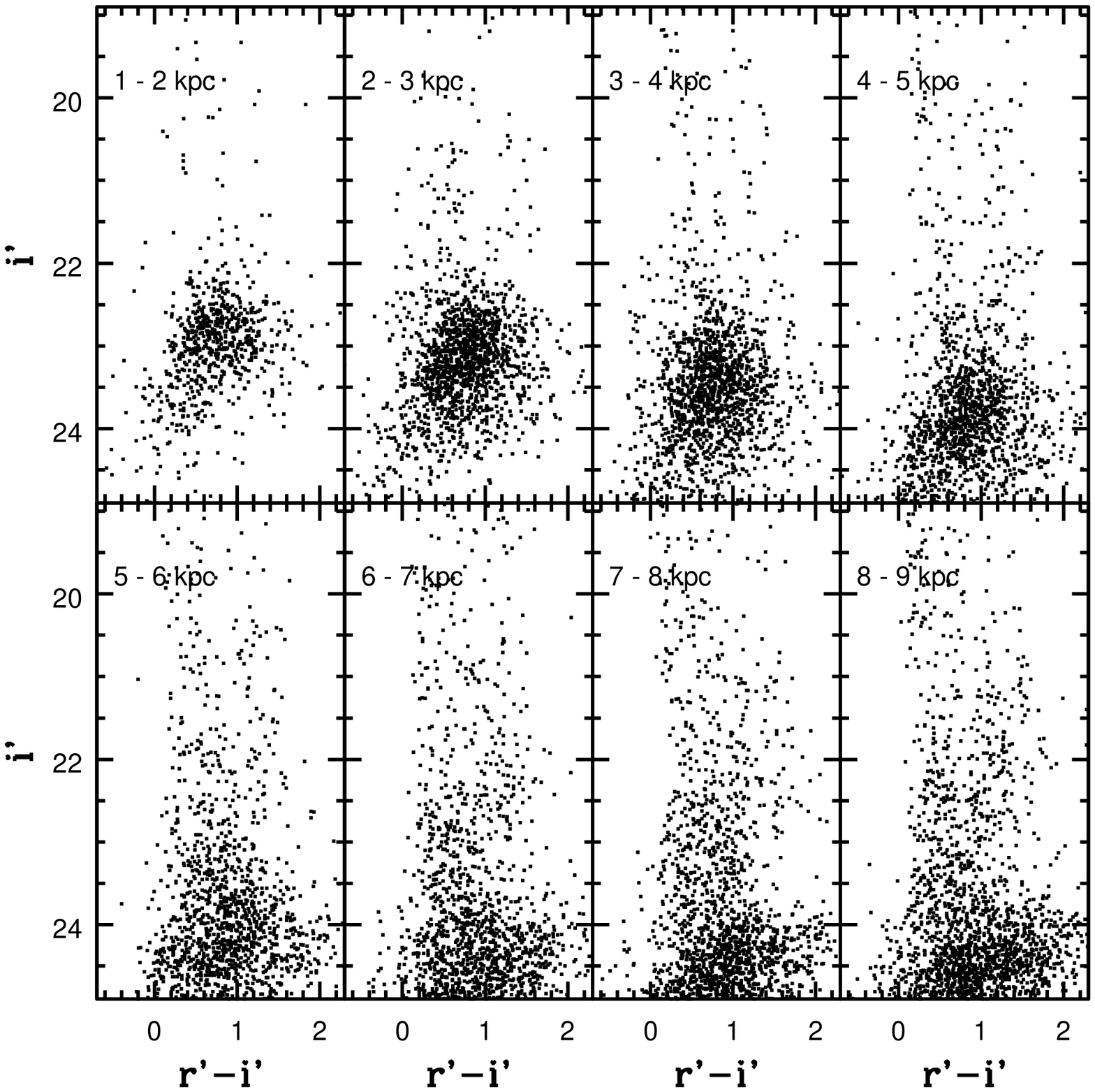}
\caption{The $(i', r'-i')$ CMDs. The distance intervals given in each panel 
are galactocentric radii, assuming a disk inclination of 64.4 degrees and a distance modulus 
of 27.5. An obvious concentration of AGB stars can be traced out to R$_{GC} \approx 7 - 
8$ kpc ($\approx 8 - 9$ disk scale lengths) in the CMDs. At larger R$_{GC}$ contamination from 
Galactic dwarfs, which form extended blue and red sequences with $i' < 22$, and background 
galaxies, which populate the diffuse collection of predominantly red objects with $i' > 22$, 
complicate efforts to detect AGB stars by eye in the CMDs.}
\end{figure}

\clearpage
\begin{figure}
\figurenum{2}
\epsscale{0.75}
\plotone{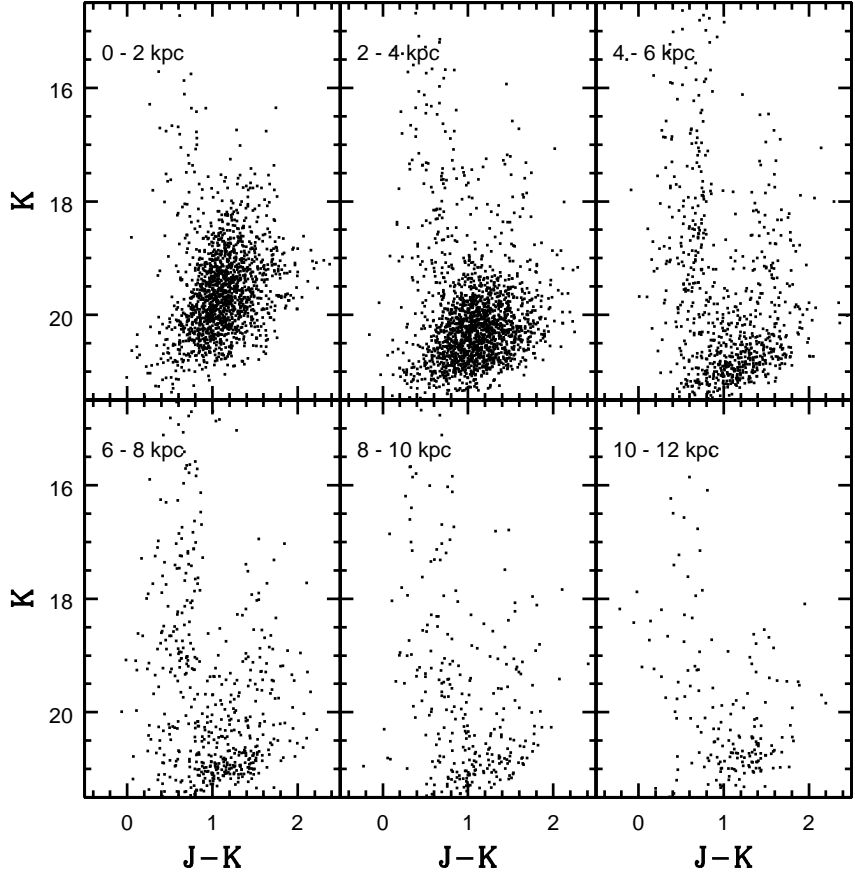}
\caption{The same as Figure 1, but showing $(K, J-K)$ CMDs. A concentration of objects 
in the CMDs that is due to AGB stars can be traced out to R$_{GC} \approx 4$ kpc 
(4.4 disk scale lengths); AGB stars are detected at larger radii in the $(i', r'-i')$ 
CMDs because of the higher number density of AGB stars on the visible/red CMDs. 
The CMDs of objects with R$_{GC} > 4$kpc are dominated by 
foreground Galactic dwarfs, which populate the sequence with $J-K \approx 0.6$, and background 
galaxies, which account for the vast majority of redder objects.}
\end{figure}

\clearpage
\begin{figure}
\figurenum{3}
\epsscale{0.75}
\plotone{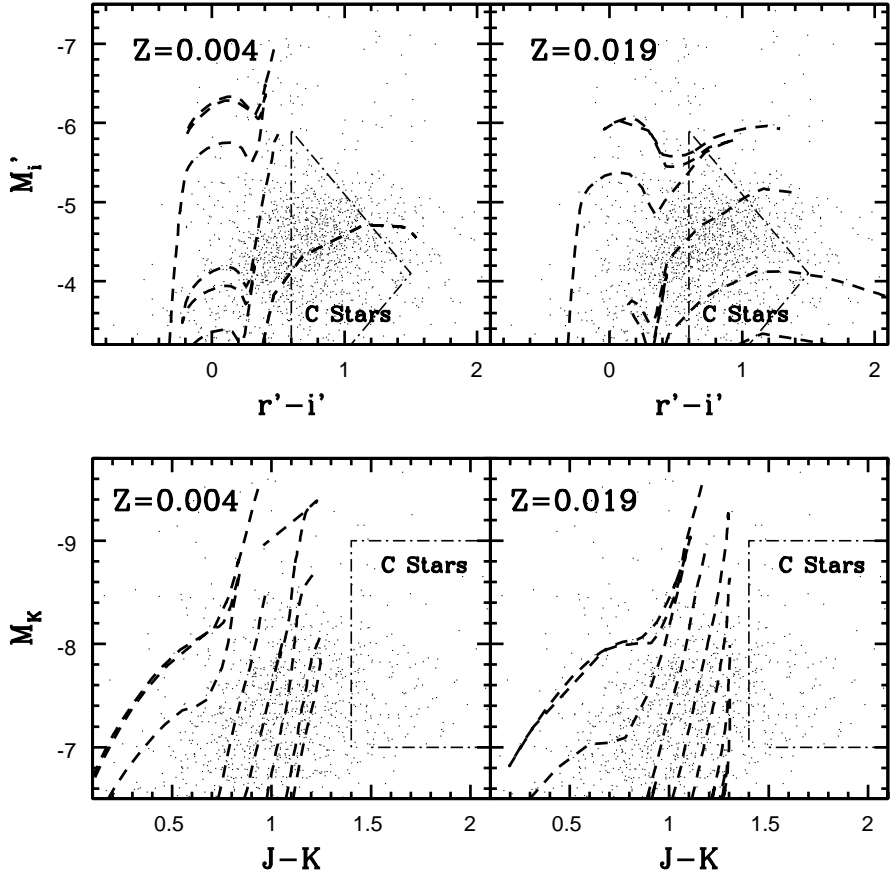}
\caption{Isochrones from Girardi et al. (2002; top panel) and Girardi et al. (2004; lower 
panel) are compared with the CMDs of stars with R$_{GC}$ between 2 and 3 kpc. 
Isochrones with log(t$_{yr}$) = 7.5, 8.0, 8.5, and 9.0, 9.5, and 10.0 are shown, 
although the last three sequences are too faint to appear on the $(i', r'-i')$ 
CMD. The region of the M31 $(i', r'-i')$ CMD that 
Demers \& Battinelli (2005) found to contain C stars 
is indicated in the upper panels, while the region of the 
LMC $(K, J-K)$ CMD from Nikolaev \& Weinberg (2000) that 
contains a distinct C star plume is indicated in the lower panel.
The blue stars on the $(i', r'-i')$ CMD, which are 
likely evolving on the MS, have log(t$_{yr}) \leq 7.5$, depending on the assumed 
metallicity. The WIRCam CMD contains stars with log(t$_{yr}) \geq 9.0$.}
\end{figure}

\clearpage
\begin{figure}
\figurenum{4}
\epsscale{0.75}
\plotone{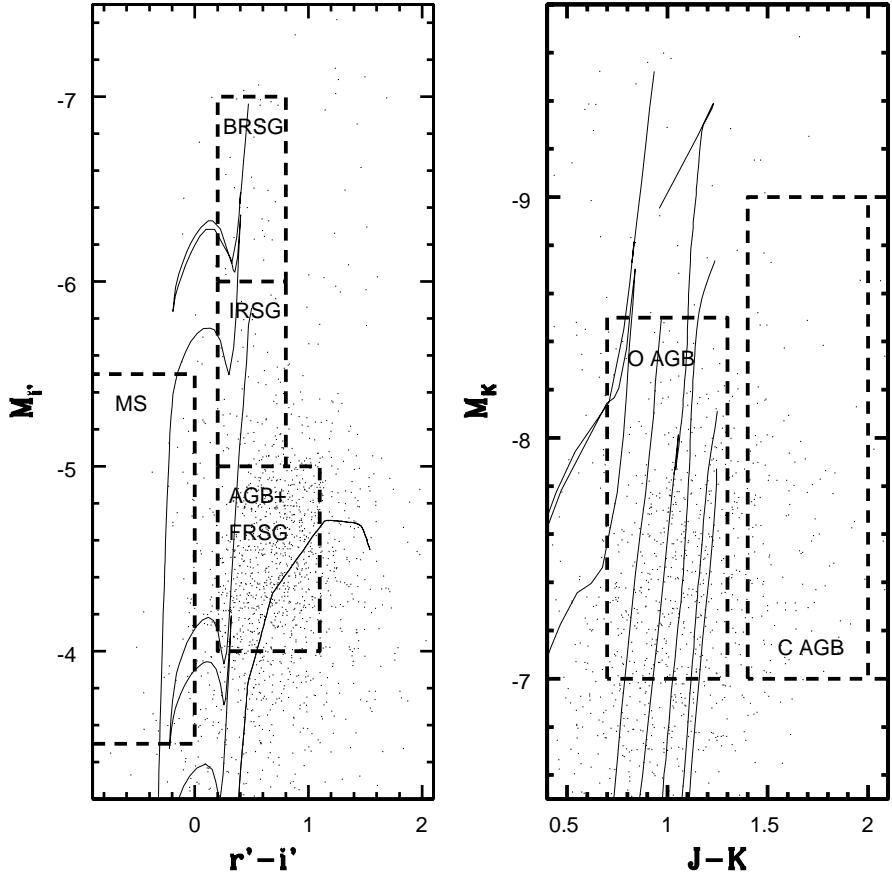}
\caption{The photometric boundaries of the stellar groups 
that are used to investigate the distribution of stars in NGC 5102. The CMDs are those of 
sources with R$_{GC}$ between 2 and 3 kpc. Isochrones with Z = 0.004 from Girardi 
et al. (2002; 2004) are shown, and these have 
ages log(t$_{yr}$) = 7.5, 8.0, and 8.5, 9.0, 9.5, and 10.0. The last 
three sequences are too faint to appear on the $(i', r'-i')$ CMD. 
Three groups of RSGs are identified on the $(i', r'-i')$ CMD: bright RSGs (BRSGs), consisting 
of stars with ages $\leq 30$ Myr, intermediate brightness RSGs (IRSGs), consisting of 
stars with ages between 30 and 100 Myr, and faint RSGs and bright AGB 
stars (FRSG/BAGBs), consisting of stars older than 100 Myr. AGB stars with oxygen (OAGB) and 
carbon (CAGB) dominated atmospheres are identified on the $(K, J-K)$ CMD. The bright and 
faint end of the CAGB box is based on the location of C stars on the LMC CMD 
discussed by Nikolaev \& Weinberg (2000).}
\end{figure}

\clearpage
\begin{figure}
\figurenum{5}
\epsscale{0.75}
\plotone{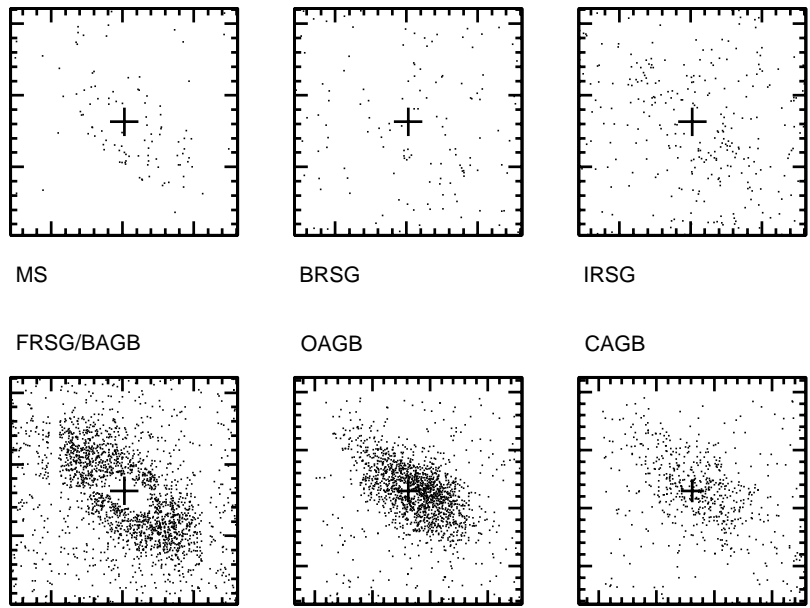}
\caption{The spatial distributions of the stellar types defined 
in Figure 4. The center of NGC 5102 is marked with a cross, and each panel is $10 
\times 10$ arcmin in size, with North at the top, and East to the left. 
MS and BRSG stars occur in largest numbers in the south-western half of the disk, 
while the IRSG and FRSG/BAGB stars are more uniformly distributed. The dearth of stars near 
the center of NGC 5102 in the FRSG/BAGB distribution is the result of crowding and gaps 
between CCDs. A concentration of OAGB stars is seen immediately to the west of the galaxy 
center, while the spatial extent of CAGB stars along the east-west axis is asymmetric, 
with CAGB stars extending to larger distances to the west than to 
the east of the center of NGC 5102. In the text it is argued that the 
concentration of OAGB stars is a companion galaxy that is projected against 
the disk of NGC 5102.}
\end{figure}

\clearpage
\begin{figure}
\figurenum{6}
\epsscale{0.75}
\plotone{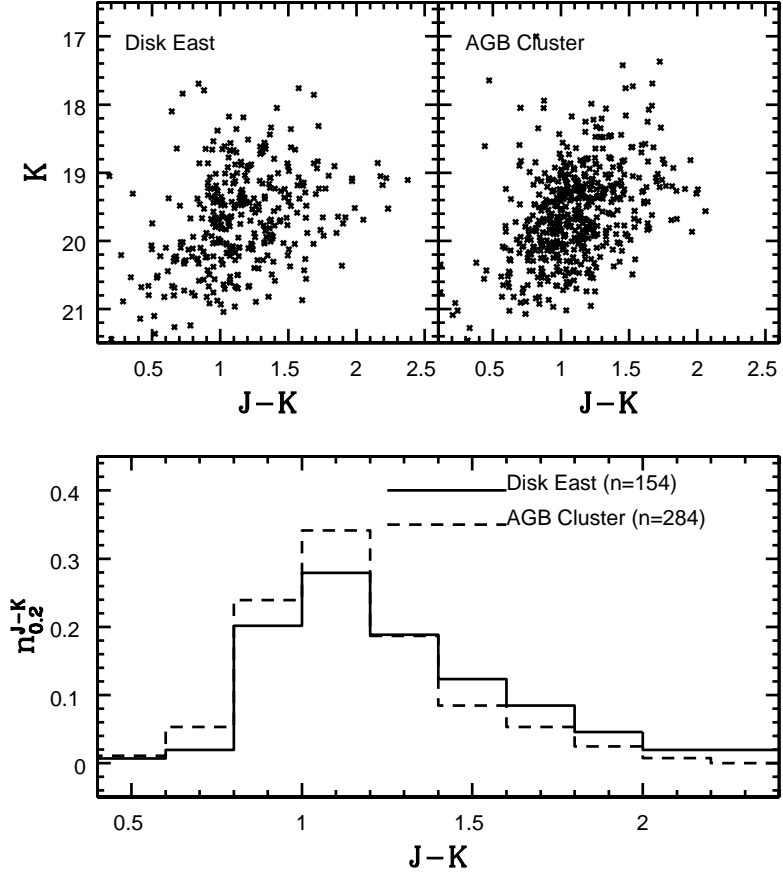}
\caption{(Top row) The $(K, J-K)$ CMDs of two 1.6 arcmin$^2$ fields that are centered 
50 arcsec to the east and west of the nucleus of NGC 5102; 
the AGB Cluster field is centered on the concentration of AGB stars 
that is seen in Figure 5. (Lower panel) The $J-K$ color distributions of 
stars with $K$ between 19 and 20 in both fields, 
where n$_{0.2}^{J-K}$ is the number of stars per 0.2 magnitude $J-K$ bin, normalized 
to the total number of stars with $J-K$ between 0.0 and 2.5. The number of stars 
that were used to construct each distribution are indicated. While both distributions 
peak near $J-K = 1.1$, the AGB Cluster distribution appears to be skewed by 
$\approx 0.2$ mag to bluer $J-K$ colors. A Kolmogorov-Smirnov test indicates that 
the two distributions differ at the 91\% significance level. A 0.2 mag difference in J--K 
corresponds to a difference in ages of at least 0.5 dex 
among stellar systems with the same metallicity.}
\end{figure}

\clearpage
\begin{figure}
\figurenum{7}
\epsscale{0.75}
\plotone{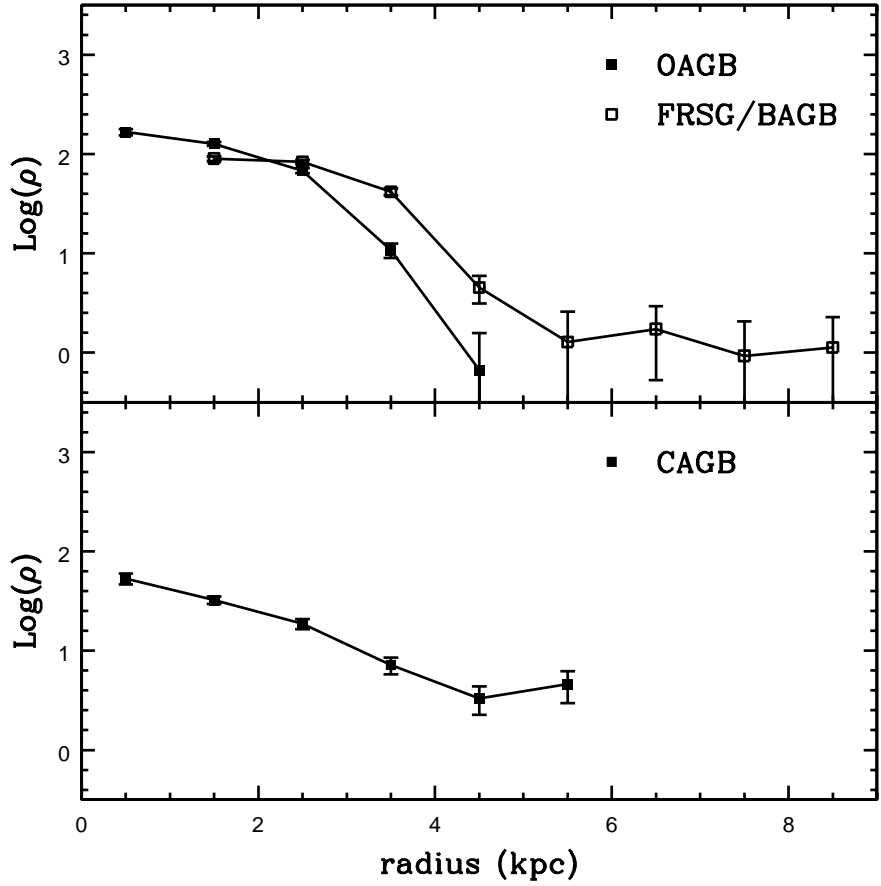}
\caption{The radial distribution of OAGB, CAGB, and FRSG/BAGB stars
in a $\pm 1$ kpc wide strip along the major axis of NGC 5102. 
$\rho$ is the number of stars arcmin$^{-2}$, corrected for background galaxies and 
foreground stars using source counts in a control field. The error bars show $1\sigma$ 
uncertainties predicted by Poisson statistics.}
\end{figure}

\clearpage
\begin{figure}
\figurenum{8}
\epsscale{0.75}
\plotone{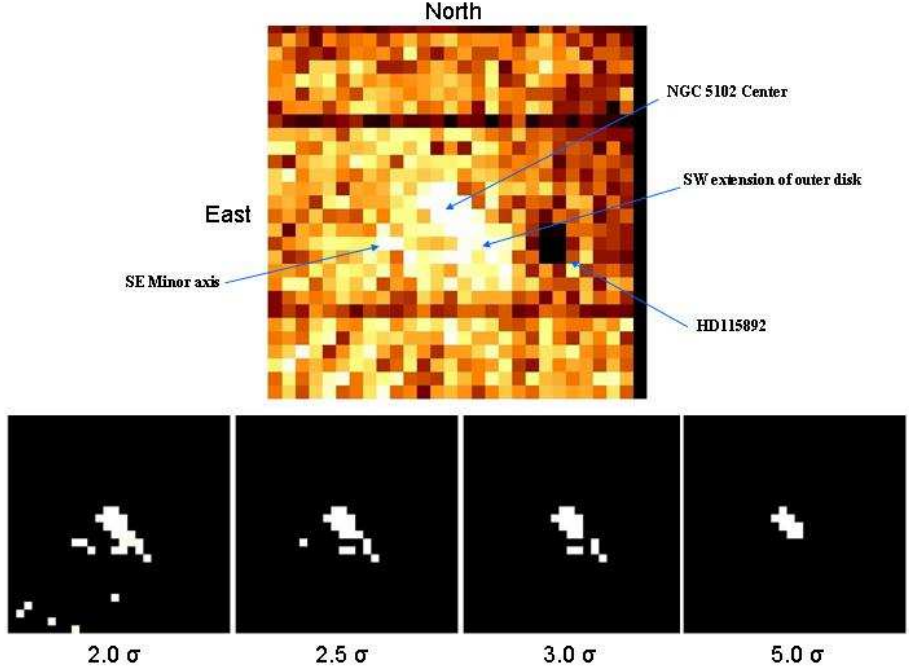}
\caption{(Top image) The distribution of FRSG/BAGB objects in the 1 degree$^2$ 
MegaCam field. Each pixel subtends $130 \times 130$ arcsec, which corresponds 
roughly to $2.1 \times 2.1$ kpc ($2.3 \times 2.3$ disk scale lengths) at the distance 
of NGC 5102. The intensity is proportional to the number of FRSG/BAGB objects in each pixel. 
The pixels with the highest numbers of FRSG/BAGBs are white, and pixels become 
progressively darker as the number of FRSG/BAGB stars drops. 
NGC 5102 is at the center of the image, and the large dark area to 
the lower right of NGC 5102 is centered on the $V = 2.8$ star HD115892 ($\iota$ Cen), 
which is a `dead zone' for detecting faint objects because of scattered light from this 
star. The two horizontal stripes are 80 arcsec gaps between CCD banks. The 
disk of NGC 5102 can be traced along the south west major axis out to a distance of 
$\approx 15$ kpc, or 16.7 disk scale lengths. Elevated star counts are also 
seen along the south east minor axis, extending out to $\approx 18$ kpc from the galaxy 
center. (Lower panels) The statistical significance of the number counts in individual 
pixels, where the pixels with number counts that exceed 
those in the control field at the 2.0, 2.5, 3.0, and 5.0 $\sigma$ levels are shown. The 
number densities of stars within 5 arcmin of the galaxy center exceed those measured in the 
control field at well above the $5\sigma$ level. Number counts in portions of the 
disk along the south west arm of the major axis exceed those in the 
control field at the $3\sigma$ level, while those along the south east arm of the minor 
axis exceed those in the control field at the $2 - 2.5\sigma$ level.}
\end{figure} 

\end{document}